# Online interactive fitting and simulation of protein circular dichroism spectra for use in education and for preliminary spectral analysis

*Luciano A. Abriata*

Laboratory for Biomolecular Modeling and Protein Production and Structure Core Facility, École Polytechnique Fédérale de Lausanne, Switzerland

luciano.abriata@epfl.ch

**Abstract** Far-UV circular dichroism (CD) spectroscopy provides a rapid, sensitive, nondestructive tool to analyze protein conformation by monitoring secondary structure composition. Originally intended for educational purposes, a spreadsheet-based program that implements a rudimentary routine for fitting and simulating far-UV protein CD spectra became quite used in research papers too, as it allowed very quick deconvolution of spectra into secondary structure compositions and easy simulation of spectra expected for defined secondary structure contents. To make such software more readily available, I present here an online version that runs directly on web browsers allowing even faster analyses on any modern device and without the need of any spreadsheet or third programs. The new version further extends the original capabilities to fit and simulate alpha, beta and random coil contents now including also beta turns; it allows to quickly select the effective spectral window used for fitting and enables interactive exploration of the effects of changes in secondary structure composition on the resulting spectra. The web app allows ubiquitous implementation in biophysics courses for example right on student smartphones, and seamless, rapid tests in research settings before moving to more advanced analysis programs (a few proposed here too). The web app is freely available without registration at http://lucianoabriata.altervista.org/jsinscience/cd/cd3.html

## Introduction

Far-UV (~190-250 nm) circular dichroism (CD) spectroscopy is highly sensitive to the secondary structure composition of proteins, providing a rapid, low-sample, nondestructive, quite widely available tool to analyze protein conformation [1–4].

Early methods for interpretation of protein CD data relied on the assumption that each type of secondary structure produces a distinctive pattern independently of its structural context. Under such assumption, given a set of basis spectra for model secondary (such as alpha helical, beta sheet and random coil) structures one can fit any experimental spectrum to the fractions of each secondary structure in the subject protein [5,6]. Although we know today that this assumption holds for some structures but breaks down for certain structural types (most notably for beta sheets, whose spectral shape depends strongly on their twisting angles [7]) the original formulation is still very useful to quickly analyze spectra, to objectively support conclusions drawn from expert experience, to teach and self-learn about the technique in structural biology courses, and to produce simulations and anticipate possible experimental results. Thus, a program that implements these simple methods as an easy-to-use MS Excel spreadsheet [8] became quite used beyond its original scope of assisting education, permeating also research [9–21]. However, such spreadsheet program is limited to users that have a copy of MS Excel with VBA enabled, and may pose usage problems for users not familiar with VBA applications. Here I present an online alternative to this program, which runs on any web browser on any laptop, tablet or smartphone, coded using standard client-side programming [22,23] that remains totally hidden from the user (contrary to the VBA code in the MS Excel spreadsheet which the user could in principle unwillingly modify). I take the opportunity to highlight the ease of use in educational settings and for rapid, preliminary investigations, and also to point readers to other modern, more robust approaches for finer analyses.

This new online tool allows fitting CD spectra to a three-state model that accounts for basic alpha, beta and random coil basis sets like the original spreadsheet, a four-state model that adds beta turns, and to a new three-state model that considers only the two beta and the random coil structures (useful to avoid interference from the strong signatures of spectra for helical structures when they are known or expected to be absent). Like the original spreadsheet program, the online version also allows the simulation of spectra corresponding to different secondary structure compositions, now more interactively as the user simply changes the contents with buttons that automatically update the simulated spectrum. Besides, the online version allows users to easily change the effective spectral window used for fitting, which may be helpful especially in situations where the low-wavelength data is noisy or suffers from distortions that lead to high detector voltages. The online web app runs on all modern web browsers in several operating systems, on computers, tablets and smartphones; and despite being an online web app, all the data analysis and processing happens in the user's computer thus not posing any privacy problems. The web app is available free of charge at http://lucianoabriata.altervista.org/jsinscience/cd/cd3.html

**Methods**
The methods behind this web app are exactly the same as in the original spreadsheet-based program [8]. Briefly, a set of basis spectra are fit analytically to minimize the least-squares error between experimental and predicted spectrum, over the specified wavelength domain. This domain must be in the range from 190 to 250 nm, over which the basis spectra are defined. By default the app proposes to use the 200-250 nm window as the region from 190 to 200 nm is often distorted in real applications.

In a simple form, the error function minimized is:

$$Err = \sum_{wl}\left(CD_{exp,wl} - \sum_{sstype} C_{sstype} BasisSpc_{sstype,wl}\right)$$

where the outer summation runs through the input wavelength data points that lie within the fitting range (defined in *Fitting Controls*), $CD_{exp,wl}$ is the CD signal at each wavelength, the inner summation runs through the number of secondary structure types analyzed (depending on which *Fit* button was clicked), $C_{sstype}$ is the fractional content of each secondary structure (ranging from 0 to 1 in the equation but displayed as percentage in the app), and $BasisSpc_{wwtype,wl}$ is the signal of each pure secondary structure at each wavelength. If the experimental spectra are provided normalized to the cell path length, number of residues and protein concentration, then the coefficients reported by the web app should already be considered final, in the range from 0 to 100%, and summing around 100%. However, if the user provides raw data then the C coefficients will all include a (same) linear constant; in such case, the *Normalize* button will re-scale all the fitted coefficients such that they sum 100% to facilitate interpretation.

Basis spectra for alpha, beta and random coil spectra are exactly the same as in the spreadsheet-based program, while a basis spectrum for beta turns was added. As in the original program, basis spectra are encoded as polynomials defined in the range [190-250] nm (function called basis_cd() in the JavaScript file sourced by the web app).

**Utilization**
To fit a spectrum, the user must (*Step 1*, Figure 1 left side) copy-paste or upload a tab- or space-separated file with a column for wavelengths followed by a column for CD signal (either as measured or normalized, see methods). Copy-pasting from any spreadsheet program works. Once the data is in the text box, clicking *Run* under *Step 2* shows the analysis, simulation and plotting controls on the right. To carry out simulations only, the user can click *Run* without loading any data.

On the tab displayed after clicking *Run* (titled *Result: experimental and simulated spectra*, Figure 1 right side) the user sees a plot of the experimental spectrum in black, in grey a simulated spectrum to be updated with the simulation and fitting controls, and reference spectra if enabled. The *Simulation controls* on the top right update the grey spectrum as expected from different basis set contributions as the user changes numbers or clicks the + and – buttons to tune the secondary structure contents. Right below, the *Normalize* button normalizes all the secondary structure contents such that they sum 100% and updates the plot (which will just be vertically shrunk or stretched but will preserve its shape). The *Download simulated data* button provides a text file with the simulated spectrum in two columns (wavelength and simulated CD signal).

Further below, the *Fitting controls* allow users to define the exact spectral window used for fitting and then buttons to fit to either three combinations of basis sets: (alpha+beta+turn+coil), alpha+beta+coil, or beta+turn+coil. The former is the most complete set, but may introduce artifacts by overfitting; the second corresponds to the earliest kinds of analysis and is the one implemented in the original spreadsheet program; the latter is useful to avoid the strong spectral dominance of the alpha helical basis set when no helical content is expected.

**Some important practical considerations**
The possibility to choose the spectral window for fitting is very important when artifacts might be present, for example from scattering and absorbance increasing too strong at low wavelengths. Fitting over different windows should return similar results if there are no big artifacts and the spectrum is correctly baselined. Important practices to optimize data collection and analysis are to avoid strong scattering and absorption, checking that the detector voltage does not saturate especially at low wavelengths, verifying that the buffer does not give a strong CD signal and to subtract it from that of the sample if present, and if possible collecting the region from 250 to 260 nm to obtain a flat baseline that can then be subtracted from the whole spectrum to remove any offset. Additional assumptions for any kind of secondary structure analysis include that the protein is pure and that it has no chromophores that absorb in the far-UV region.

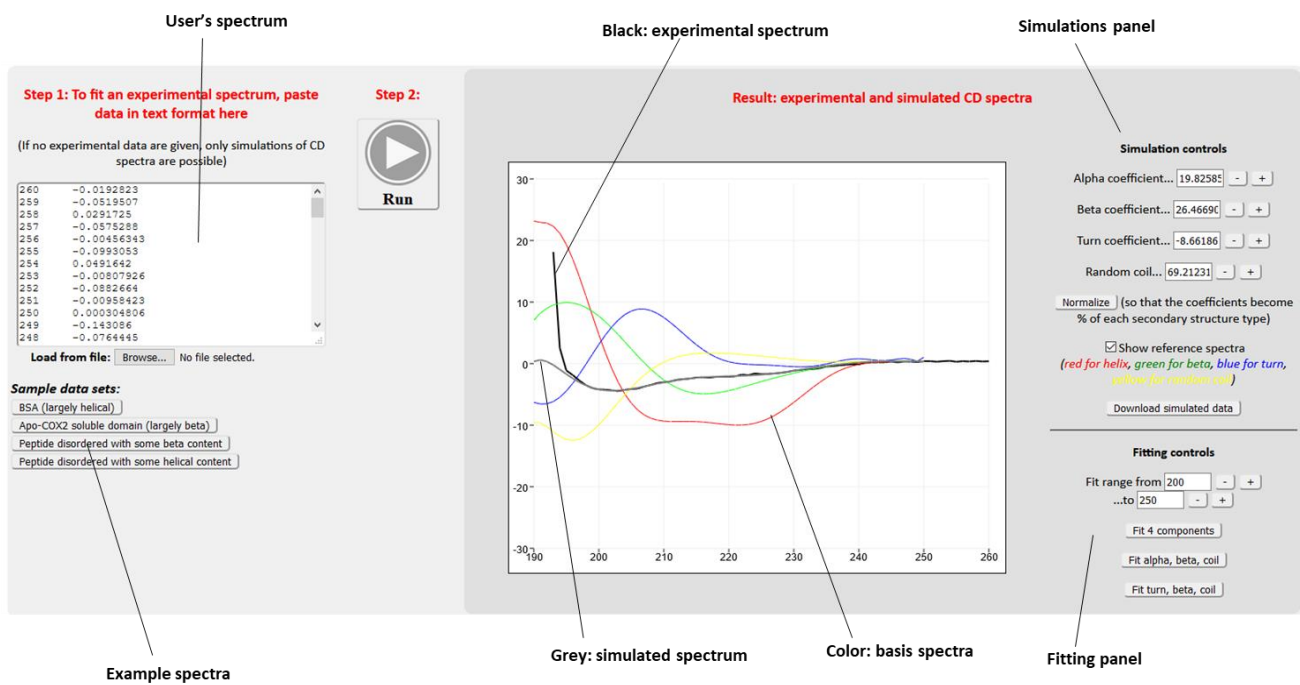

**Figure 1.** The full web app interface (after having clicked "Run" under Step 2), here displaying the spectrum from the example titled "Peptide disordered with some helical content" in black and the result of fitting to 4 components in grey. Freely available at http://lucianoabriata.altervista.org/jsinscience/cd/cd3.html

A final, important consideration, is that as explained in the introduction the methods used by this web app and the original spreadsheet program are very simple and of an early generation that has now been superseded by more complex approaches. They are good enough for education and in research for quick secondary structure estimations (with up to 10-20% inaccuracies in the structure contents), for simulating spectra and predicting the effects that changes in secondary structure composition could have on spectra. For finer analyses of far-UV CD spectra programs less interactive but that use more modern approaches and datasets are recommended. For example, BeStSel [24] considers several secondary structure types including different types of helices and different beta twists for CD data fit. For simulations, PDB2CD [25] and DichroCalc [26] enable direct simulations from structures, the latter also including near-UV circular dichroism and linear dichroism.